# Understanding topological phase transition in monolayer transition metal dichalcogenides

Duk-Hyun Choe,[*] Ha-Jun Sung, and K. J. Chang[†]

*Department of Physics, Korea Advanced Institute of Science and Technology, Daejeon 34141, Rep. of Korea*

[*]E-mail: (D.-H.C.) hoalad@gmail.com. [†]E-mail: (K.J.C.) kjchang@kaist.ac.kr

Abstract

Despite considerable interest in layered transition metal dichalcogenides (TMDs), such as $MX_2$ with M = (Mo, W) and X = (S, Se, Te), the physical origin of their topological nature is still poorly understood. In the conventional view of topological phase transition (TPT), the non-trivial topology of electron bands in TMDs is caused by the band inversion between metal $d$ and chalcogen $p$ orbital bands, where the former is pulled down below the latter. Here, we show that, in TMDs, the TPT is entirely different from the conventional speculation. In particular, $MS_2$ and $MSe_2$ exhibits the opposite behavior of TPT, such that the chalcogen $p$ orbital band moves down below the metal $d$ orbital band. More interestingly, in $MTe_2$, the band inversion occurs between the metal $d$ orbital bands. Our findings cast doubts on the common view of TPT and provide clear guidelines for understanding the topological nature in new topological materials to be discovered.

## I. INTRODUCTION

The non-trivial topology of electron bands in solids gives rise to many intriguing physical phenomena. Notable examples include topological (crystalline) insulators [1-5], which exhibit a bulk insulating gap as well as gapless edge/surface states that are topologically protected from scattering, and Weyl semimetals [6-7], which are characterized by the gapless bulk states with Fermi arcs on the surface. In such materials, the so-called topological materials, the basic mechanism behind the non-trivial band topology is the band inversion between two orbital bands, which can be driven by chemical bonding [8-10], spin-orbit coupling (SOC) [11-13], and applying pressure [3,14-18] or electric field [19]. The understanding of the band inversion in topological materials is crucial, because it not only provides the most intuitive way to explain the topological phase transition but also determines the helicities of the spin textures of the surface states [12,20]. As shown schematically in Fig. 1(a), the inverted band structure, together with the insulating bulk gap formed by inter-band hybridization, is commonly adopted to describe the energy dispersion [9-13,18,21,22] in topological insulators (TIs). This feature has naturally led to the speculation that the topological phase can be determined by analyzing the orbital characters of the electronic states near the Fermi level.

Recently, monolayer transition metal dichalcogenides (TMDs) in the 1T' (or distorted octahedral) phase have been predicted to be two-dimensional (2D) TIs [23], and both their bulk and few-layered forms have been synthesized on a large scale with high quality [24,25]. These layered materials have attracted increasing interest owing to their potential applications to novel electronic and spintronic devices based on quantum spin Hall effect [1,23] and their possible candidates for the realization of Majorana fermions [26]. However, understanding of their topological nature is still in its infancy. It was proposed that a spontaneous structural phase transition from the 1T (or octahedral) to 1T' phase [Fig. 1(b)] causes the band inversion and thereby leads to the topological phase transition from trivial to non-trivial [23,25]. As the energy dispersion of 1T'-MoS$_2$ strongly resembles the common feature for topological materials [Fig. 1(c)], the band inversion in 1T'-TMDs is therefore described [23,27,28], based on the common speculation that the chalcogenide $p_x$ band, which usually forms the valence band, moves above the metal $d_{xz}$ band (henceforth referred to as the *d-p* inversion mechanism). In contrast, however, our finding is that the common feature is clearly not the case; instead, counter-intuitively, the $p_x$ and $d_{xz}$ bands are already inverted even before the structural phase transition takes place. On the other hand, the band topology of Te-based TMDs, such as MoTe$_2$, exhibits an entirely different feature as compared to MoS$_2$ [23,25]. It is therefore essential to provide a general feature for understanding the underlying mechanism behind such peculiar band topologies in TMDs.

In this work, we investigate the topological phase transition in monolayer TMDs, where we find band inversion mechanisms entirely different from the conventional speculation. This is done by developing a scheme for analyzing the complex band structure of topological materials. Based on our scheme, we show that 1T'-MX$_2$ exhibits two distinct types of band inversion. In 1T'-MS$_2$ and 1T'-MSe$_2$, completely opposite to the conventional speculation, the X-*p* band is pulled down below the M-*d* band (henceforth referred to as the *p-d* type band inversion). More interestingly, in 1T'-MTe$_2$, the two different M-*d* bands are inverted during the topological phase transition, which is here called the *d-d* type band inversion. Our findings shed light on the precise mechanism behind topological nature topological materials, offering new insights into the underlying processes.

## II. CALCULATION METHOD

Our first-principles calculations were performed within the framework of density functional theory, which employed the generalized gradient approximation (GGA) for the exchange-correlation potential [29] and the projector augmented wave (PAW) potentials [30], as implemented in the VASP code [31]. The wave functions were expanded in plane waves up to an energy cutoff of 600 eV. We used a repeated supercell geometry with a vacuum region larger than 15 Å to prohibit interactions between adjacent supercells. We chose Γ-centered $k$-points generated by 20×10 Monkhorst-Pack meshes for Brillouin zone integration, and used the GGA optimized lattice parameters (see Table I).

## III. RESULTS AND DISCUSSION

Group-VI layered TMDs ($MX_2$, where M = Mo, W, X = S, Se, Te) exhibit a wide range of intriguing electronic properties, depending on their crystal structures [32]. Among various polytypes in $MX_2$, the most well-studied and stable polytype is a trigonal prismatic 2H phase, which possesses a moderate bandgap (> 1.0 eV) and exhibits strong valley polarization due to inversion symmetry breaking, whereas an octahedral 1T phase is metallic, with inversion symmetry. The 1T phase, although it can be engineered by carrier doping [33] or phase patterning [34], is known to be a metastable phase due to its intrinsic dynamical instability. Driven by electron-phonon interactions [25], the 1T phase spontaneously transforms to the more stable 1T' phase, which also has inversion symmetry, exhibiting Peierls-like distortions, as shown in Fig. 1(b).

When the 1T phase undergoes a structural transformation to the 1T' phase in monolayer $MX_2$, the band topology changes from trivial to non-trivial. To elucidate the topological phase transition in details, we consider several intermediate structures ranging from the 1T to 1T' phase. The intermediate structures, referred to as 40%-1T', 60%-1T', and 80%-1T', were obtained by linearly interpolating the atomic positions and lattice parameters between the 1T and 1T' phases and then optimizing only the X atoms with fixing the M atoms at the interpolated positions. The same supercell geometry [Fig. 1(b)] is used for all the considered structures to directly compare their atomic and electronic properties. In general, the topology of the time-reversal-invariant band structure is characterized by the $Z_2$ invariant $v$, in which $v = 0$ and $v = 1$ indicate the trivial and non-trivial states, respectively. Since the lattice inversion symmetry is present in all the intermediate structures as well as the 1T and 1T' phases, the $Z_2$ invariant can be calculated by the product of the parity eigenvalues[3,4] at the time-reversal-invariant momenta (TRIM), Γ, X, Y, and S [25]. For all the TMDs considered, we note that the topological phase transition from $v = 0$ to $v = 1$ is solely attributed to the band inversion at the Γ point, whereas the product of the parity eigenvalues does not change at the other three TRIM points (see Table II).

In $MX_2$, however, it is difficult to identify the band inversion mechanism simply by analyzing the shape of the band structure because of the significant modification of the electronic structure [25]. To clearly illustrate the band inversion process near the Γ point, we introduce a different concept of the orbital-parity (OP) projected band structure. The Kohn-Sham wave function for the $n$th band at the wave vector $k$ is written as $|\psi_{k,n}\rangle = \sum_{i,l,m} c_{k,n}^{i,l,m} |i,l,m\rangle$, where $|i,l,m\rangle$ is the PAW projector for the $i$th atom, $l$ is the angular momentum, and $m$ is the magnetic quantum number. The complex coefficient $c_{k,n}^{i,l,m}$ can be decomposed as $c_{k,n}^{i,l,m} = |c_{k,n}^{i,l,m}| e^{i\phi_{k,n}^{i,l,m}}$, where $\left|c_{k,n}^{i,l,m}\right|$ is the orbital weight and $e^{i\phi_{k,n}^{i,l,m}}$ is the phase factor. When the inversion center is chosen as the origin, the phase factors at the Γ point should satisfy the relation, $e^{i\phi_{\Gamma,n}^{A,l,m}} = \pm e^{i\phi_{\Gamma,n}^{A',l,m}}$, due to the inversion symmetry, where $A$ and $A'$ denote an inversion pair of the M or X atoms in MX$_2$. Now we define the effective parity index such as $p_{k,n}^{A,l,m} = (-1)^l \cos\left(\phi_{k,n}^{A,l,m} - \phi_{k,n}^{A',l,m}\right)$, which ranges from −1 (odd parity) to +1 (even parity). This index measures the degree of symmetry for each wave function and its values are equivalent to the parity eigenvalues at the Γ point. We further define the orbital-parity (OP) weight such as $w_{k,n}^{A,l,m} = \left|c_{k,n}^{A,l,m}\right|^2 p_{k,n}^{A,l,m}$. For any centrosymmetric materials, $w_{k,n}^{A,l,m}$ allows us to clearly identify both the orbital characters and the parities of the wave functions in the band structure. The characterization of both orbital and parity is well visualized for 1T'-MoS$_2$, which is taken as an example (see Fig. 2).

Figure 3(a) illustrates the evolution of the OP projected band structure around the Γ point during the 1T-to-1T' transition in MoS$_2$. We first focus on the S-$p_x$ orbital band with odd parity and the Mo-$d_{xz}$ orbital band with even parity near the Fermi level, which are represented as $\left|\text{S-}p_x^-\right\rangle$ and $\left|\text{Mo-}d_{xz}^+\right\rangle$, respectively. In the 1T' phase, the $\left|\text{S-}p_x^-\right\rangle$ and $\left|\text{Mo-}d_{xz}^+\right\rangle$ bands exhibit the inverted-like energy dispersion by the $d$-$p$ inversion [Figs. 1(a) and 1(c)], with the suggestive inverted gap of $\Delta = 0.55$ eV. However, as apparently shown in Fig. 3(a), the band ordering between the $\left|\text{S-}p_x^-\right\rangle$ and $\left|\text{Mo-}d_{xz}^+\right\rangle$ states remains unchanged throughout the whole structural transformation. Contrary to the previous speculation [23], the $\left|\text{S-}p_x^-\right\rangle$ and $\left|\text{Mo-}d_{xz}^+\right\rangle$ states are not involved in the band inversion. This is rather counter-intuitive as the two inverted-like bands near the Fermi level do not play any role in the topological phase transition. Similarly, we find that the $\left|\text{X-}p_x^-\right\rangle$ state always lies above the $\left|\text{M-}d_{xz}^+\right\rangle$ state in both the 1T and 1T' phases of MX$_2$ (Fig. 4), indicating that these two states are irrelevant to the topology change for all the group-VI TMDs. With including the SOC, the band gap opens in 1T'-MX$_2$ [23,25], however, the SOC is not sufficiently strong enough to reverse the band order near the Fermi level, as shown in Fig. 3.

Now we discuss three orbital bands that play a crucial role in the topological phase transition, the M-$d_{yz}$ orbital band with even parity ($|\text{M-}d_{yz}^+\rangle$), the X-$p_y$ orbital band with odd parity ($|\text{X-}p_y^-\rangle$), and the M-$d_{z^2}$ orbital band with odd parity ($|\text{M-}d_{z^2}^-\rangle$). The evolution of the OP projected band structure is shown for MoS$_2$ and MoTe$_2$ in Fig. 3. In MoS$_2$, we find that the topological phase transition follows a different mechanism, the $p$-$d$ type band inversion, in which the $|\text{S-}p_y^-\rangle$ state moves down below the $|\text{Mo-}d_{yz}^+\rangle$ state during the evolution. In 60%-1T' MoS$_2$, it is clear that the $|\text{S-}p_y^-\rangle$ and $|\text{Mo-}d_{yz}^+\rangle$ bands cross near the Fermi level, as shown in the inset of Fig. 3(a). After that, these two bands move apart from each other, with a significant modification of their band shape, and eventually form a large inverted gap of $\delta_{\text{MoS}_2}$ = 1.05 eV [Fig. 3(a)]. Note that $\delta_{\text{MoS}_2}$ must be distinguished from the gap $\Delta$ in Fig. 1(c), because $\Delta$, which will be derived from the band crossing between the $|\text{S-}p_x^-\rangle$ and $|\text{Mo-}d_{xz}^+\rangle$ states, already exists in the 1T phase. On the other hand, a different characteristic feature is found in MoTe$_2$. In MoTe$_2$, the $|\text{Mo-}d_{z^2}^-\rangle$ band is pulled down below the $|\text{Mo-}d_{yz}^+\rangle$ band [Fig. 3(b)], resulting in an inverted gap of $\delta_{\text{MoTe}_2}$ = 0.94 eV. This $d$-$d$ band inversion is the fundamental mechanism for the topology change of MoTe$_2$, in contrast with the MoS$_2$ case. As will be discussed shortly, the above two band inversion mechanisms can explain the topological phase transition in all the group-VI TMDs.

For better understanding of the band inversion mechanism in MX$_2$, the *p-d* and *d-d* inversion processes are schematically drawn in Fig. 5(a). The $|\text{M-}d^+_{xz,yz}\rangle$, $|\text{X-}p^-_{x,y}\rangle$, and $|\text{M-}d^-_{z^2}\rangle$ states mainly contribute to the energy states around the Fermi level during the structural transformation from 1T to 1T'. In the 1T phase, owing to the octahedral crystal fields with trigonal distortion (D$_{3d}$ group symmetry), the $|\text{M-}d^+_{xz}\rangle$ and $|\text{M-}d^+_{yz}\rangle$ states are degenerate, and the $|\text{X-}p^-_x\rangle$ and $|\text{X-}p^-_y\rangle$ states are also degenerate. As the metal M atoms deviate from their octahedral positions [Fig. 1(b)], the degeneracies of the *d*- and *p*-orbital bands are fully lifted. The energy splitting of the $|\text{M-}d^+_{xz,yz}\rangle$ states and their evolution during the structural transformation are similar for both the *p-d* and *d-d* inversion processes, as shown in Fig. 5(a). The marked difference between the two processes stems from the band order of the $|\text{X-}p^-_{x,y}\rangle$ and $|\text{M-}d^-_{z^2}\rangle$ states in the 1T phase. The calculated energies of the $|\text{M-}d^+_{xz,yz}\rangle$, $|\text{X-}p^-_{x,y}\rangle$, and $|\text{M-}d^-_{z^2}\rangle$ states are compared for 1T-MX$_2$ with X = S, Se, Te, and Po in Fig. 5(b). For a pedagogical reason, we consider MPo$_2$, in which Po has six valence electrons ($6s^2 6p^4$) as in S, Se, and Te. We find that the chalcogenide atom plays a crucial role in determining the band ordering in 1T-MX$_2$ and thereby the type of band inversion. As shown in Fig. 5(c), as the atomic number of X increases, the $|\text{X-}p^-_{x,y}\rangle$ levels increase, simply because the valence electrons loosely bound to the nucleus. In 1T-MS$_2$ and 1T-MSe$_2$, the $|\text{X-}p^-_{x,y}\rangle$ states are positioned below the $|\text{M-}d^-_{z^2}\rangle$ state. As a consequence, the energy splitting of the $|\text{X-}p^-_{x,y}\rangle$ states leads to the level crossing between the $|\text{X-}p^-_y\rangle$ and $|\text{M-}d^+_{yz}\rangle$ bands, i.e., the *p-d* band inversion. On the other hand, in 1T-MoTe$_2$ and 1T-MPo$_2$, the $|\text{X-}p^-_{x,y}\rangle$ states are higher in energy, hence, the splitting of the $|\text{X-}p^-_{x,y}\rangle$ states does not affect the band topology, as shown in Fig. 5(a). Instead, the $|\text{M-}d_{z^2}^-\rangle$ band is pulled down below the Fermi level, eventually leading to the level crossing between the $|\text{M-}d_{z^2}^-\rangle$ and $|\text{M-}d_{yz}^+\rangle$ bands, i.e., the *d-d* band inversion. When the $|\text{M-}d_{z^2}^-\rangle$ and $|\text{X-}p_{x,y}^-\rangle$ bands are close to each other, especially for 1T-WTe$_2$ and 1T-WPo$_2$, these two orbital bands are strongly hybridized during the structural transformation. Thus, the band inversion is given by a mixture of the *p-d* and *d-d* inversions.

Finally, we performed hybrid functional (HSE06) calculations [35] to describe more accurately the band structures of MX$_2$. It is known that the GGA calculations underestimate the conduction bands of semiconductor and insulators, thus, incorrect band orderings may occur in MX$_2$. For the HSE06 calculations, we used the screening parameter set to $\omega = 0.207$ Å$^{-1}$ and chose the lattice parameters and atomic positions optimized by GGA. For all the TMDs considered, we find that the band order of the $|\text{M-}d_{xz,yz}^+\rangle$, $|\text{X-}p_{x,y}^-\rangle$, and $|\text{M-}d_{z^2}^-\rangle$ states is not affected [see Fig. 5(c)], confirming that the *p-d* and *d-d* inversion mechanisms play a role in the topological phase of MX$_2$.

## IV. CONCLUSION

By taking layered TMDs as an example, we have shown that the band inversion mechanism in topological materials cannot be directly inferred from their band structures even though it is intuitively obvious. This indicates that the conventional view of topological phase transitions may lead to incorrect conclusions, casting doubt on the previously known band inversion mechanisms. Moreover, our OP projection scheme would provide guidelines for understanding the topological phase in new topological materials to be discovered. Recently, it was reported that, in three-dimensional TIs, such as NaBaBi [12] and Hg-based chalcogenides [20], the left- or right-handed spin textures of the surface states are determined by the type of band inversion. The effect of the two different inversion mechanisms in TMDs on their topological edge states will be an interesting subject for future studies. In addition, owing to the two $d$ orbital bands involved in the topological phase transition, Te-based TMDs may also exhibit electron-electron correlation effect [12,36]. Thanks to recent advances in the synthesis of monolayer 1T- and 1T'-TMDs [34,37,38], it may soon be possible to explore these effects experimentally.

## ACKNOWLEDGMENT

D.-H. Choe thanks M. Tomić for his technical help. This work was supported by Samsung Science and Technology Foundation under Grant No. SSTF-BA1401-08.

**TABLE I.** The GGA optimized lattice constants (in units of Å) of 1T- and 1T'-MX$_2$.

|  | 1T phase | | 1T' phase | |
|:---:|:---:|:---:|:---:|:---:|
|  | *a* | *b* | *a* | *b* |
| MoS$_2$ | 3.18 | 5.51 | 3.17 | 5.72 |
| MoSe$_2$ | 3.28 | 5.68 | 3.28 | 5.96 |
| MoTe$_2$ | 3.50 | 6.06 | 3.44 | 6.38 |
| WS$_2$ | 3.20 | 5.54 | 3.20 | 5.72 |
| WSe$_2$ | 3.29 | 5.70 | 3.30 | 5.95 |
| WTe$_2$ | 3.51 | 6.08 | 3.50 | 6.31 |

**TABLE II.** For monolayer 1T- and 1T'-MX$_2$, the Z$_2$ invariant $v$ is calculated by analyzing the parities of the wave functions at the time reversal invariant momenta (TRIMs) in the Brillouin zone, where $(-1)^v = \prod_{k=\Gamma,X,Y,S} \delta_k$. At the four TRIMs ($\Gamma$, X, Y, and S), $\delta_k$ is given by the product of the parity eigenvalues, $\delta_k = \prod_{m=1}^{N} \xi_{2m}(k)$. In the 1T phase, all the Z$_2$ invariants are 0, indicating the trivial band topology for 1T-MX$_2$. On the other hand, in the 1T' phase, all the Z$_2$ invariants are 1, indicating the non-trivial band topology for 1T'-MX$_2$. The topological phase transition from $v = 0$ to $v = 1$ in MX$_2$ is solely attributed to the band inversion at the $\Gamma$ point, whereas $\delta_k$ does not change at the other three TRIMs.

|  |  | $\delta_{k=\Gamma}$ | $\delta_{k=X}$ | $\delta_{k=Y}$ | $\delta_{k=S}$ | $v$ |
|---|---|---|---|---|---|---|
| 1T phase | MoS$_2$ | − | − | − | − | 0 |
|  | MoSe$_2$ | − | − | − | − | 0 |
|  | MoTe$_2$ | − | − | − | − | 0 |
|  | WS$_2$ | − | − | − | − | 0 |
|  | WSe$_2$ | − | − | − | − | 0 |
|  | WTe$_2$ | − | − | − | − | 0 |
| 1T' phase | MoS$_2$ | + | − | − | − | 1 |
|  | MoSe$_2$ | + | − | − | − | 1 |
|  | MoTe$_2$ | + | − | − | − | 1 |
|  | WS$_2$ | + | − | − | − | 1 |
|  | WSe$_2$ | + | − | − | − | 1 |
|  | WTe$_2$ | + | − | − | − | 1 |

**FIGURES**

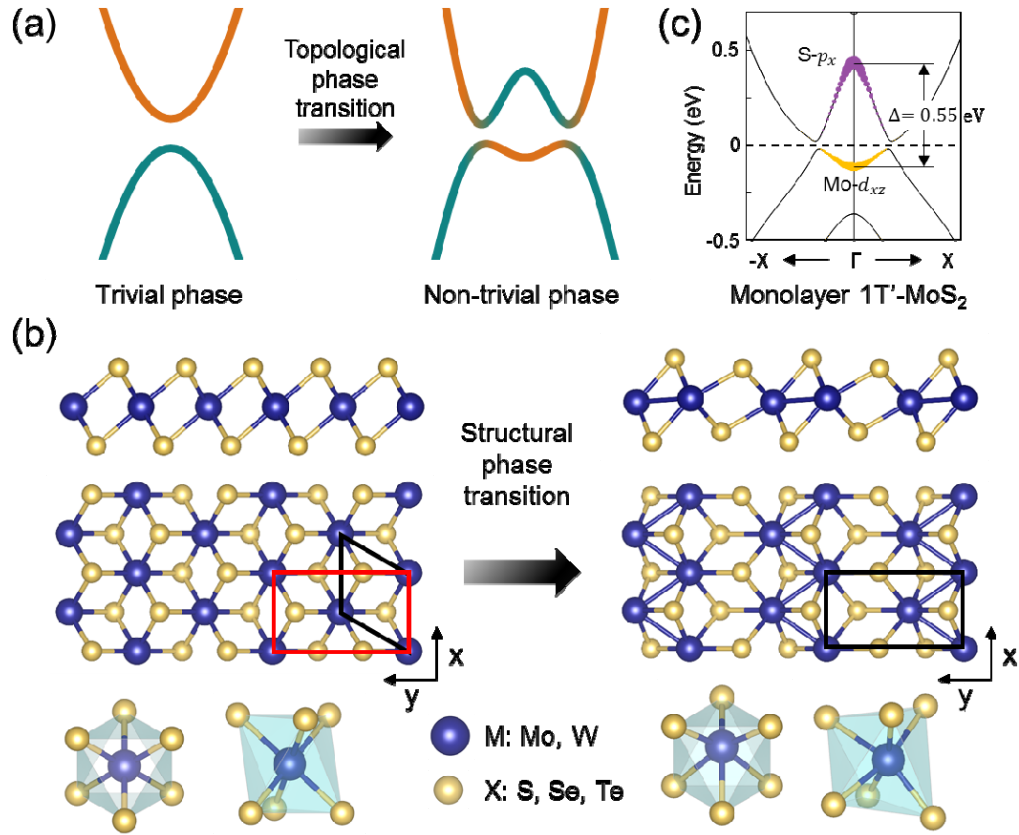

**FIG. 1.** (a) A schematic picture for the topological phase transition in TIs. The valence band rises above the conduction band, and inter-band hybridization opens the non-trivial energy gap. Based on this principle, the band inversion mechanism can be directly inferred from the band structure. (b) The structural transformation from the 1T to 1T' phase in TMDs. The primitive cells are denoted as black solid lines, whereas red solid lines indicate the rectangular supercell for the 1T phase. (c) The band structure of monolayer 1T'-MoS2 near the Γ point strongly resembles the schematic picture in (a), exhibiting a suggestive inverted gap of $\Delta = 0.55$ eV and thus misleading the band inversion mechanism in 1T'-TMDs.

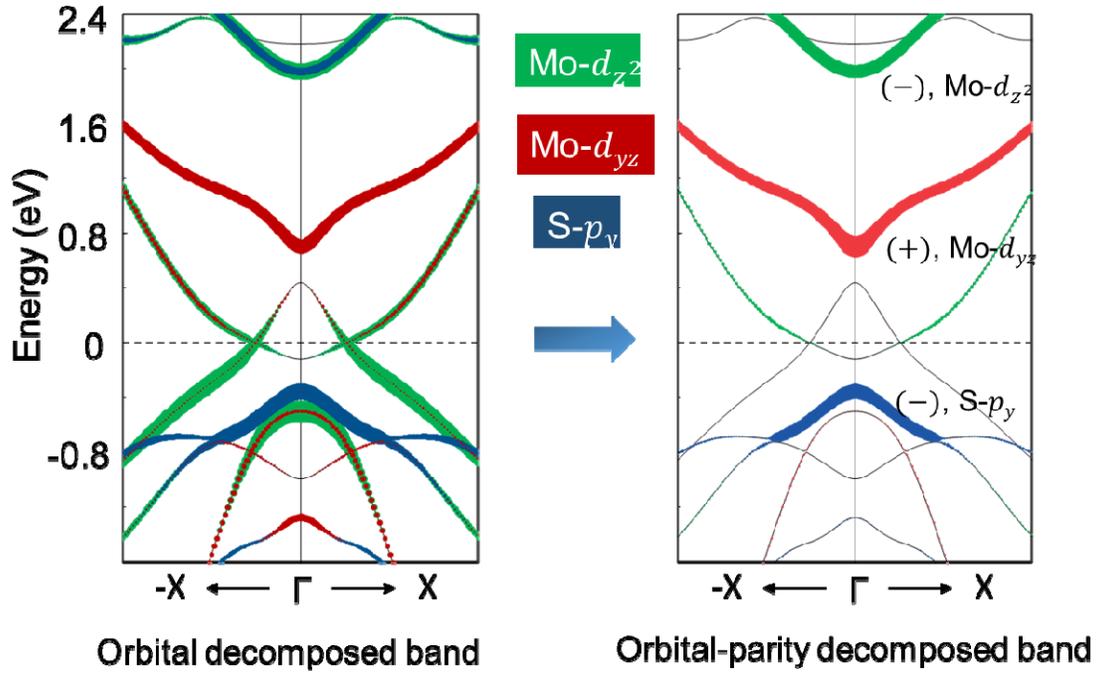

**FIG. 2.** Comparison of the orbital decomposed band structure with the orbital-parity (OP) decomposed band structure in 1T'-MoS$_2$. Red, blue, and green dots denote the Mo-$d_{yz}$, S-$p_y$, and Mo-$d_{z^2}$ orbital states, respectively. In the OP decomposed band structure, both the orbital character and the effective parity of the wave function are clearly visualized.

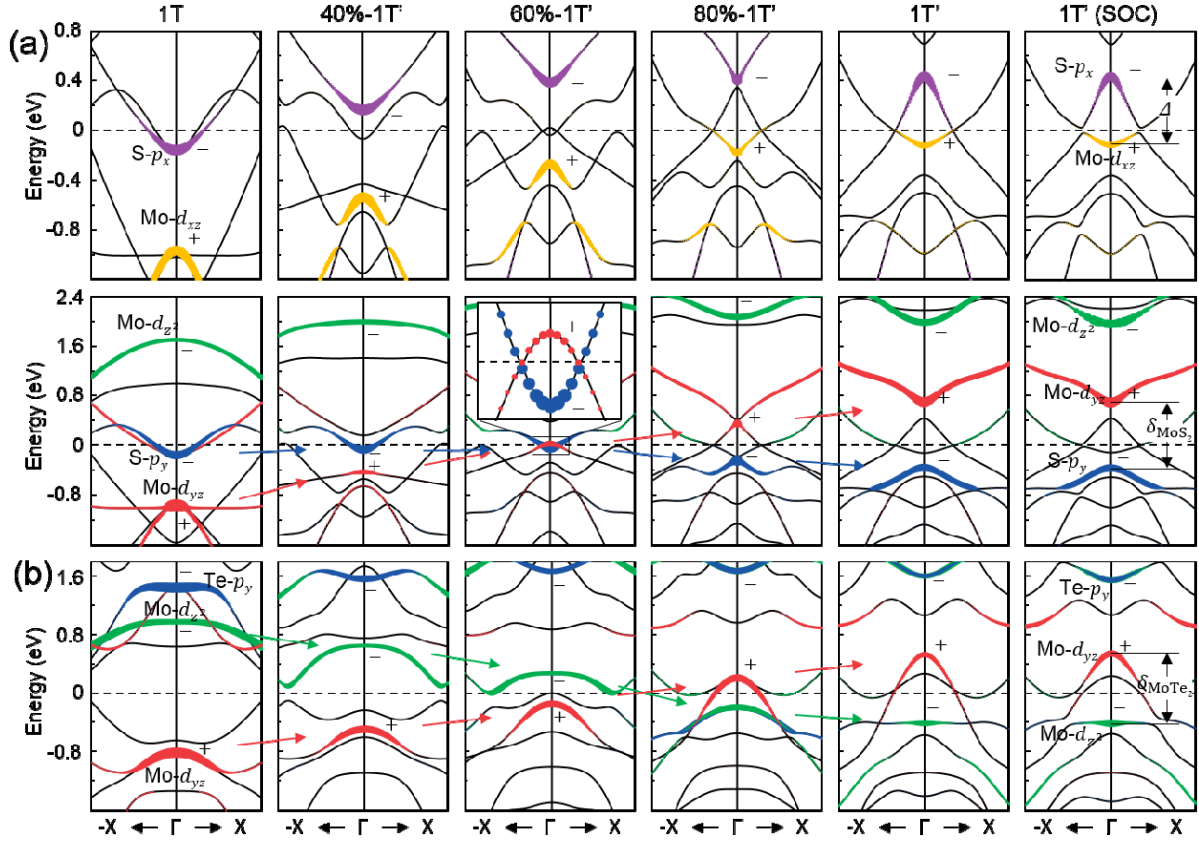

**FIG. 3.** The evolution of the band structures during the structural transformation from 1T to 1T' in $MX_2$. (a) The orbital-parity (OP) decomposed band structures of (a) $MoS_2$ and (b) $MoTe_2$. Purple and yellow dots indicate the $|S\text{-}p_x^-\rangle$ and $|Mo\text{-}d_{xz}^+\rangle$ states, respectively. Red, blue, and green dots indicate the $|M\text{-}d_{yz}^+\rangle$, $|X\text{-}p_y^-\rangle$, and $|M\text{-}d_{z^2}^-\rangle$ states, respectively. The rightmost panels show the band structures in the 1T' phase with including the spin-orbit coupling.

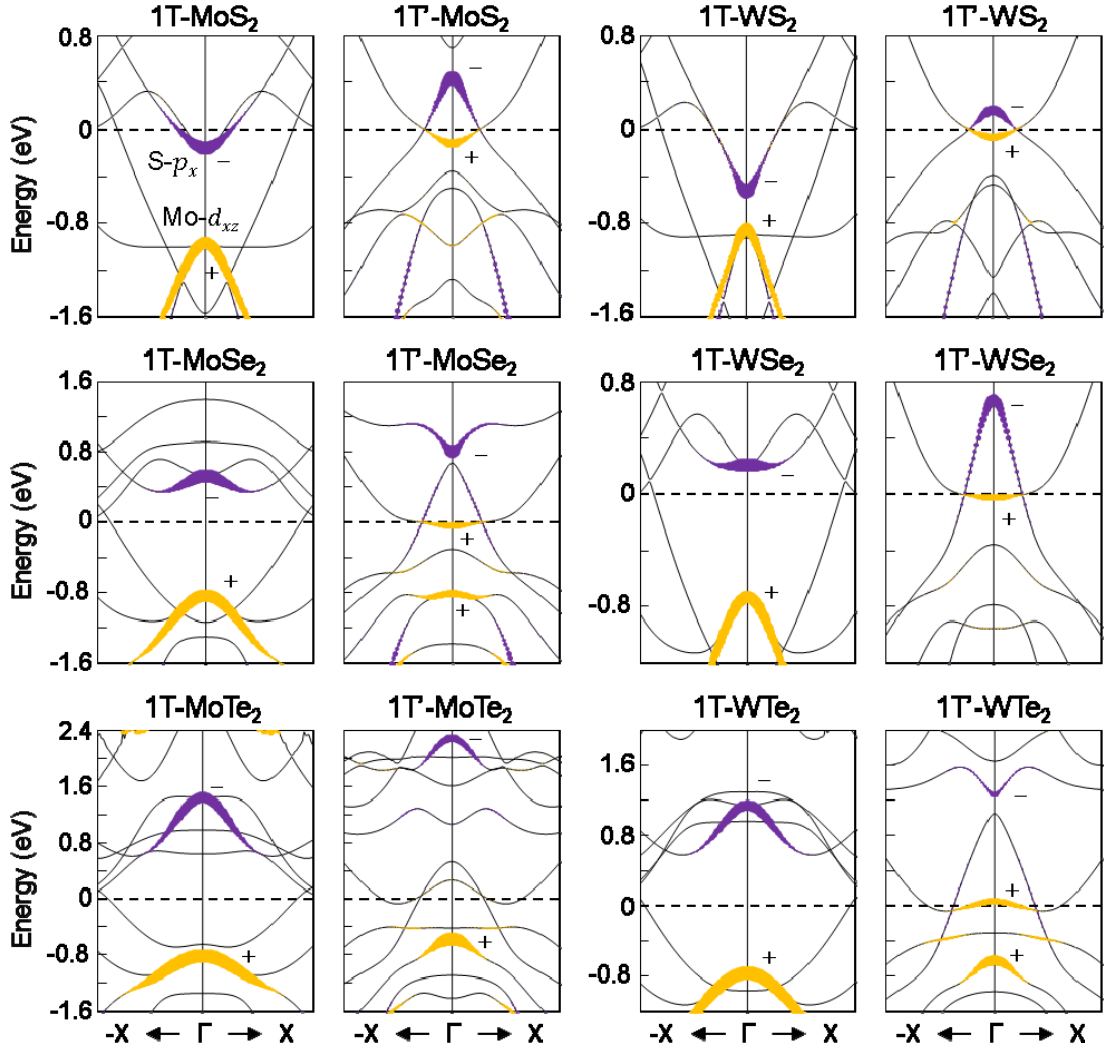

**FIG. 4.** The OP decomposed band structures of group-VI TMDs in the 1T and 1T' phases without SOC. Purple and yellow dots denote the $|\text{X-}p_x^-\rangle$ and $|\text{M-}d_{xz}^+\rangle$ states, respectively, and their corresponding parities are labelled. For all the TMDs, the $|\text{X-}p_x^-\rangle$ state lie above the $|\text{M-}d_{xz}^+\rangle$ state, indicating that these states are not involved in the band inversion process.

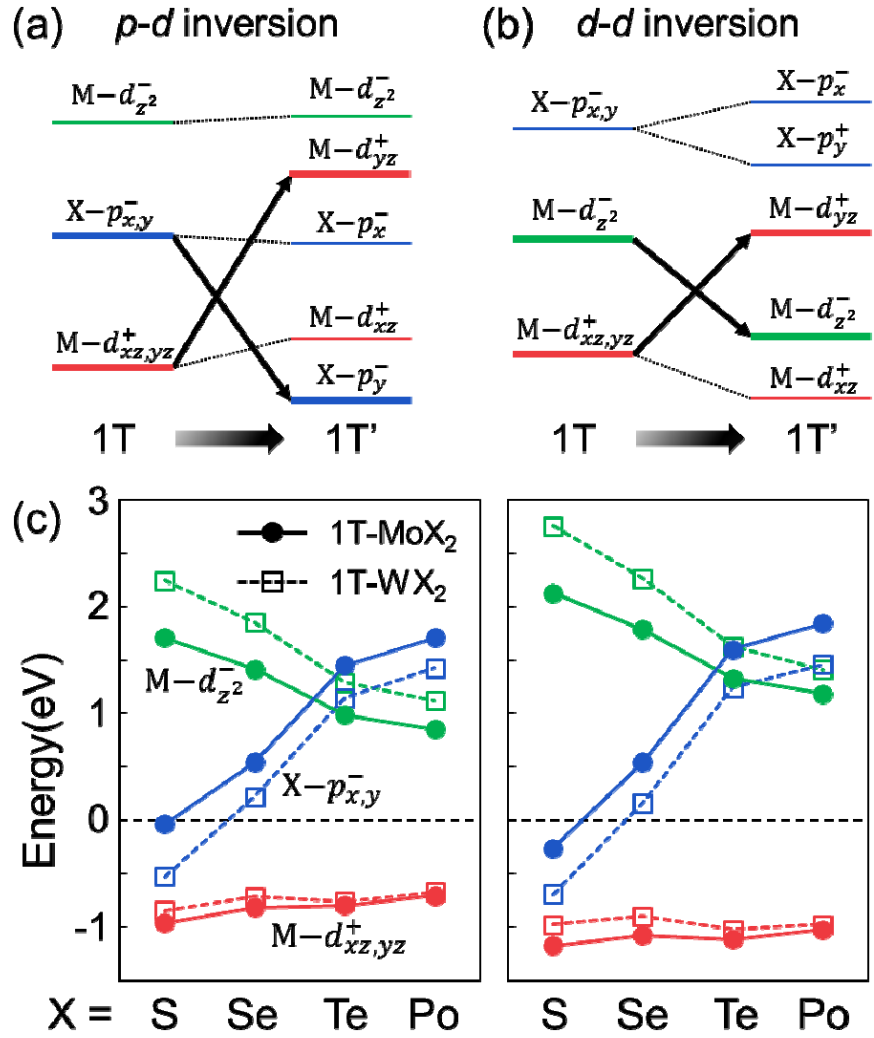

**FIG. 5.** Schematic diagrams for the (a) *p-d* and (b) *d-d* band inversion processes in MX$_2$. (c) The GGA (left panel) and HSE06 (right panel) calculations for the band ordering in 1T-MX$_2$.